# Location Intelligence Reveals the Extent, Timing, and Spatial Variation of Hurricane Preparedness

Bo Li[1*], and Ali Mostafavi[2]

[1] (Corresponding author) Ph.D. Student, Urban Resilience.AI Lab, Zachry Department of Civil and Environmental Engineering, Texas A&M University, College Station, United States; e-mail: libo@tamu.edu

[2] Associate Professor, Urban Resilience.AI Lab Zachry Department of Civil and Environmental Engineering, Texas A&M University, College Station, United States; e-mail: amostafavi@civil.tamu.edu

**Abstract:** Hurricanes are one of the most catastrophic natural hazards faced by residents of the United States. Improving the public's hurricane preparedness is essential to reduce the impact and disruption of hurricanes on households. Inherent in traditional methods for quantifying and monitoring hurricane preparedness are significant lags, which hinder effective monitoring of residents' preparedness in advance of an impending hurricane. This study establishes a methodological framework to quantify the extent, timing, and spatial variation of hurricane preparedness at the census block group level using high-resolution location intelligence data. Anonymized cell phone data on visits to points-of-interest for each census block group in Harris County before 2017 Hurricane Harvey were used to examine residents' hurricane preparedness. Four categories of points-of-interest, grocery stores, gas stations, pharmacies and home improvement stores, were identified as they have close relationship with hurricane preparedness, and the daily number of visits from each CBG to these four categories of POIs were calculated during preparation period. Two metrics, extent of preparedness and proactivity, were calculated based on the daily visit percentage change compared to the baseline period. The results show that peak visits to pharmacies often occurred in the early stage of preparation, whereas the peak of visits to gas stations happened closer to hurricane landfall. The spatial and temporal patterns of visits to grocery stores and home improvement stores were quite similar. However, correlation analysis demonstrates that extent of preparedness and proactivity are independent of each other. Combined with synchronous evacuation data, CBGs in Harris County were divided into four clusters in terms of extent of preparedness and evacuation rate. The clusters with low preparedness and low evacuation rate were identified as hotspots of vulnerability for shelter-in-place households that would need urgent attention during response. Hence, the research findings provide a new data-driven approach to quantify and monitor the extent, timing, and spatial variations of hurricane preparedness. Accordingly, the study advances data-driven understanding of human protective actions during disasters. The study outcomes also provide emergency response managers and public officials with novel data-driven insights to more proactively monitor residents' disaster preparedness, making it possible to identify under-prepared areas and better allocate resources in a timely manner.

1. **Introduction**

Hurricanes are the most deadly and destructive among all recorded weather hazards in US history, according to National Oceanic and Atmospheric Administration. From 1980 to 2020, hurricanes have caused $945.9 billion in damages and 6593 deaths in the United States. (NOAA). Impacts and disruptions are typically less severe for residents who take steps to prepare in advance of hurricane landfall. (Mayer, Moss, & Dale, 2008). Preparedness actions prior to hurricane landfall would typically involve stocking up on supplies of food, water, and medicine; performing home retrofits, and procuring backup resources to increase the household's ability to cope with hurricane impacts. Inadequate preparedness leaves shelter-in-place households without needed food, water, and life supplies and would increase the hardship and well-being impacts on them (Coleman, Esmalian, & Mostafavi, 2020; Dargin, Berk, & Mostafavi, 2020). Despite its importance in hurricane response, hurricane preparedness may vary across affected areas. Therefore, insight into the extent, timing, and spatial variation of hurricane preparedness is essential knowledge to help emergency managers allocate resources and facilitate timely response to vulnerable populations.

Surveys are the most-widely used approach for examining households' hurricane preparedness. For example, Baker (2011) conducted telephone interviews with 1200 Florida households about their levels of preparedness during hurricane seasons. Josephson, Schrank, and Marshall (2017) investigated hurricane preparedness of small businesses by telephone interviews over a ten-month period. Despite its popularity, the use of surveys to investigate hurricane preparedness presents four critical limitations. First, surveys are resource- and time-consuming. Conducting telephone interviews or mailed or emailed surveys usually takes several months because the procedure evolves through a series of steps: selecting eligible respondents, contacting respondents, and waiting for responses. Time and cost make it impossible to monitor preparedness before an impending hurricane in a proactive manner. Hence, insights obtained from surveys suffer from significant lags that constrain their use to inform disaster response decisions and actions. Second, the sample size of surveys usually is not too large. The average low response rate makes the number of actual responses even smaller. Studies using survey could only use sampling methods to acquire understanding of the population instead of getting large scale data of interested area directly. Third, information collected from interviews and surveys are based on the perceptions, opinions or memories of respondents, which might vary from the actual actions taken by the respondents. Fourth, data collection about hurricane preparedness using surveys put a burden on affected communities. Hence, an alternative approach is critical to better and more proactively monitor the extent, timing, and the spatial variation of hurricane preparedness.

The limitations of surveys could be overcome with the increasing availability and popularity of location-based big data. Location intelligence data, also known as location-based service data, could provide important insights regarding human activities and mobility both during normal times, as well as in times of crises ((Fan, Jiang, & Mostafavi, 2021; Gao et al., 2020; Lee, Chou, & Mostafavi, 2022; Lee, Maron, & Mostafavi, 2021). Location intelligence data is usually collected passively from smartphone devices through GPS, Wi-Fi, and Bluetooth (Darzi, Frias-Martinez, Ghader, Younes, & Zhang, 2021). Data collected in this way is usually large-scale, free of observer error, and nearly real-time, which provides new opportunities for understanding complex dynamics of human activities and mobility.

Location intelligence data has shown promise in disaster research. Researchers have used large-scale geo-based data to analyze population displacement pattern after the 2010 Haiti earthquake (Lu, Bengtsson, & Holme, 2012); developed methods to characterize flood impacts on a population (Pastor-Escuredo et al., 2014); simulated and predicted human movement pattern after disaster (Song et al., 2016); detected impacts of extreme events on human movement (Roy, Cebrian, & Hasan, 2019); assessed flood inundation status (farahmand, Wang, Mostafavi, & Maron, 2021); quantified resilience based on the magnitude of impacts and time-to-recovery (Hong, Bonczak, Gupta, & Kontokosta, 2021); evaluated hurricane perturbation on urban mobility (Fan et al., 2021); assessed short-term disaster recovery by evacuation return and home switch (Lee et al., 2022). Nevertheless, the majority of studies have harnessed location-based data for examining human mobility and evacuation in post-disaster contexts, and the potential of these data for examining pre-disaster human preparedness activities has not been realized.

In the pre-disaster stage, several studies have focused on evacuation issues. For example, Darzi et al. (2021) utilized mobile phone location-based service data to explore evacuation patterns during Hurricane Irma. Deng et al. (2021) combined high-resolution location-based human mobility data and social-demographical information to reveal race and wealth disparities in disaster evacuation patterns. Another important pre-disaster human activity is preparedness, a critical protective action enabling shelter-in-place households to better cope with disaster disruptions and impacts. As observed in recent hurricanes in the United States, with the more rapidly accelerating hurricanes, residents lack adequate time to evacuate and rather tend to shelter in place. This tendency increases the significance of preparedness and the importance of proactive evaluation of preparedness extent, timing, and spatial variation to inform public officials and emergency managers and responders.

To fill the gap, this study adopted location intelligence data to reveal the extent, timing, and spatial patterns of hurricane preparedness. In this paper, hurricane preparedness is examined from the perspective of residents' protective action prior to disasters: visits to critical facilities to procure food, water, medicine, and gasoline in anticipation of disruptions and impacts. With high-resolution location intelligence data, this study aimed to establish a new approach to investigating and monitoring hurricane preparedness by detecting and analyzing visits from home census block groups (CBG) to critical facilities to uncover the extent, timing, and spatial variations of preparedness at the CBG level. We investigated hurricane preparedness in the context of the 2017 Hurricane Harvey in Harris County, Texas. As shown in Fig. 1, this study characterized hurricane preparedness based on three dimensions: (1) extent of preparedness, (2) proactivity of preparedness, and (3) spatial distribution. Accordingly, the study is designed around the following research questions: (1) What is the extent of hurricane preparedness of each CBG based on metrics of point-of-interest (POI) visit fluctuation? (2) How early do people start to prepare for the hurricane? (3) What are spatial hotspots of susceptibility considering hurricane preparedness and evacuation rate? To address the research questions, three indicators related to examining hurricane preparedness are defined and quantified based on location intelligence data related to residents' visits from their home CBG to points of interest which are critical to hurricane preparation.

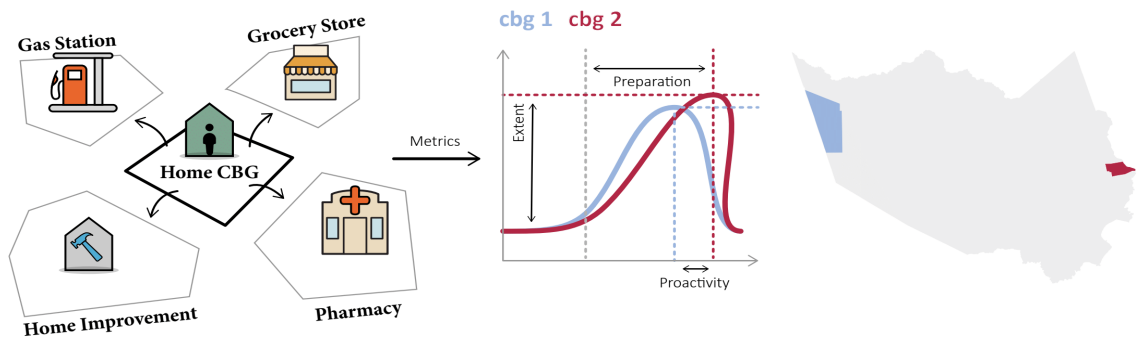

**Fig. 1**. Delineating hurricane preparedness extent and timing using location intelligence data

The remainder of the paper is organized as follows. First, datasets used in the study are described. Next, the methods for developing the study are presented in a step-by-step manner. Then, we present the empirical results with large-scale location intelligence data collected from the 2017 Hurricane Harvey in Harris County, Texas. Finally, the results are discussed, and the conclusion remarks are given in the final sections.

## 2. Data description

### 2.1 Study Context

We collected data from the 2017 Hurricane Harvey in Harris County, Texas to examine hurricane preparedness of the residents. Hurricane Harvey was a devastating Category 4 hurricane, which hit Texas and Louisiana severely in August 2017. It brought the most significant tropical cyclone rainfall ever recorded in US history both in scope and peak rainfall amounts (NOAA). and caused significant flooding damage. Harris County, which includes the Houston metropolitan area, is one of the areas most adversely affected by Hurricane Harvey.

### 2.2 Data sources

The location intelligence data used in the study was obtained from Spectus (formerly known as Cuebiq), a mobility data platform. Spectus provides high-quality location data sets by partnering with smartphone apps to collect data from devices whose users opted in to location data collection. Spectus builds its geo-behavioral dataset by cooperating with app developers to gather high-resolution dataset by Bluetooth, GPS, WiFi, and IoT signals. For each anonymous user, more than a hundred of data points on average are collected each day, providing an opportunity to gain more accurate and precise knowledge of human mobility. Currently, the scale of data collected by Spectus is roughly 15 million daily active users in the United States. High standards of privacy policy are adopted to enable ethical and responsible data collection and use. All data is collected transparently after consent, and users are free to opt out of location sharing at any time. All data Spectus provides is de-identified to ensure it is anonymous, while also undergoing additional privacy enhancements, such as the removal of sensitive POIs and the obfuscation of home areas at the census block group level. Through its Social Impact program, Spectus provides mobility insights for academic research and humanitarian initiatives. A study performed by Wang, Wang, Cao, Chen, and Ban (2019) compared Spectus data with cellular network and in-vehicle GPS data and came to the conclusion that Spectus data outperforms other sourced data by a superior combination of large scale, high accuracy, precision, and observational frequency. Beyond these benefits, Spectus data is demonstrated as highly demographically representative (Aleta et al., 2020; Nande et al., 2021; Wang et al., 2019).

In addition to providing device-level location-based data, Spectus also aggregates data using artificial intelligence and machine learning techniques. Spectus' responsible data sharing framework enables us to query anonymized, aggregated, and privacy-enhanced data by providing access to an auditable and on-premise sandbox environment. In this study, we used one of the Spectus aggregated datasets, the Device_Location table to identify the home CBGs of devices. The Device_Location table provides information at the device level, including timestamp, privacy-compliant device ID, and geo-information. We adopted dwell time of devices, which refers to the duration of the visit, as an indicator for detecting home CBG. Table Stop from core data assets was used to extract user POI visits. Based on the aggregated datasets, we calculated the number of visits, which are introduced in Methods session, to understand hurricane preparedness.

The second dataset is Microsoft building footprint. This dataset is generated and released by Microsoft free of charge. In the dataset, more than 120 million high-quality building footprints in United States are detected from satellite imagery relying on the Open Source CNTK Unified Toolkit. The building footprint data was used to determine the location of points-of-interest for data processing of human mobility data to determine patterns of visits to POIs.

The third dataset is from SafeGraph, which sources POI data by crawling open store locators on the web, using publicly available API, and crawling open web domains with updated locations for a specific category of POIs, processing and modeling to infer additional attributes, and licensing third-parties to fill in the gaps. Accuracy of the data is verified by combing machine learning techniques with human capital. We used datasets from SafeGraph to identify the location of POIs. The Core Places, a sub-dataset of SafeGraph, is a comprehensive dataset of POIs for places globally, containing location information and geographic coordinates, brand and business attributes and North American Industry Classification System (NAICS) and categorical coding. This data enables identifying specific POI types visited during hurricane preparedness. Socio-demographic data was retrieved from the American Community Survey database administrated by US Census Bureau at the CBG level. The study adopted mainly median household income to understand whether income affects the extent of hurricane preparedness. The data is the 2017 estimates over the 2015-2019 period.

## 3. Methods

Fig. 2 shows the methodological framework illustrating data processing and metrics calculation procedures.

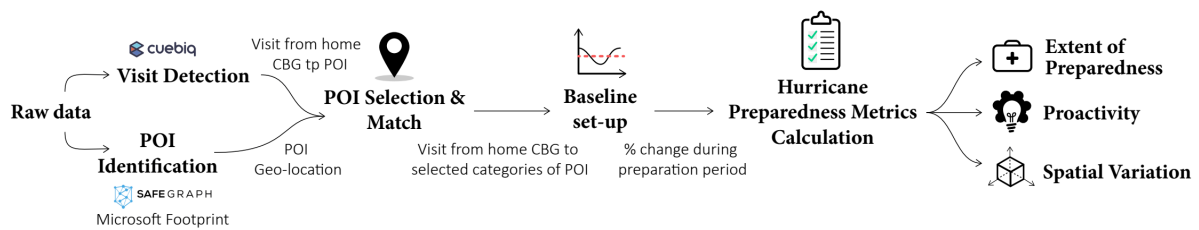

**Fig. 2.** Data processing and analysis steps.

## 3.1 Detecting visits from home CBG to POI

The first step of data processing was to use Spectus data to detect visits from home CBG to POI. One of the Spectus aggregated datasets, Device_Location table from core data assets, was used to identify home CBG of devices. The Device_Location table provides information at the device level, including timestamp, privacy compliant device ID, geo-information. We adopted dwell time of devices, which refers to the duration of the visit, as an indicator for detecting home CBG. If a device has stopped in a CBG for more than one day, the CBG will be recognized as the home CBG of the device.

Table Stop from core data assets was used to extract which POIs the users have visited. Device stop is defined as a point in space where a device spent some time. It was computed from the granular device location points using a clustering algorithm based on spatio-temporal proximity. Similarly, the indicator dwell time records the stop duration of devices and was adopted to decide whether it is a visit of the POI or just a stop-by. If a piece of data is defined as a visit to a POI, the latitude and longitude of the POI will be recorded.

From Microsoft footprints dataset, we specified the polygons of POIs. Building footprint polygon geometries were acquired in GeoJson format. Comparing centroid of polygons and the latitude and longitude of POI acquired from Spectus data, the data from two sources were matched and merged. The next step was to link brand information with POI polygons, so that one could know from the data the classification of a visited POI. NAICS is developed by Federal statistical agencies to classify business establishments, such as residential building construction, educational services, grocery wholesalers. Using NAICS code could help divide POIs into different categories and provide insight on the role of the POI in people's daily life. As mentioned before, SafeGraph incorporates a dataset, the Core Places, which includes geographic information, brands and corresponding NAICS code. By matching polygon information between footprints datasets and SafeGraph data, POI data is merged with the brand information and NAICS code.

Through the above-mentioned steps, visits from home CBG to POI with NAICS code was obtained. The original data was hourly and was temporally aggregated to daily. The indicator and descriptions of the data are shown in Table 1.

**Table 1**
Data description

| Data | Description |
| --- | --- |
| Home_CBG | The ID of home CBG |
| Place_id | The ID of POI |
| Count | The number of daily visits |
| NAICS Code | NAICS category of the POI |
| Date | Date of the visits |

**3.2 POI visits during hurricane preparedness**

In the next step, we selected the most relevant categories of POI for hurricane preparedness. In making preparations for an impending hurricane, not only the direct impact of hurricane itself must be considered, but also its cascading effects. The direct effects of hurricanes are heavy wind and rainfall. Wind could induce structural damage to lifeline infrastructure, such as electric power delivery systems (Reed, Powell, & Westerman, 2010), water supply systems (Der Sarkissian, Cariolet, Diab, & Vuillet, 2022), and telecommunication power (Kwasinski, Weaver, Chapman, & Krein, 2009). Rainfall may cause flooding and inundate transportation systems (Zhu, Hu, & Collins, 2020). As a result, people in hurricane-impacted areas may encounter a series of adverse events, such as power outage (Reed et al., 2010), lack of medical service (Ruskin et al., 2018), fuel shortage (Comes & Van de Walle, 2014), and low food security (Clay & Ross, 2020). Since people may lose access to critical facilities during and after hurricanes, procuring supplies before the disaster could largely strengthen people's capacity to cope with hurricane-induced perturbations and impacts. From the literature, four categories of POIs were identified to be the most relevant critical facilities for hurricane preparedness: grocery store, pharmacy, gas station and home improvement store. The selection was made based on the criticality of supplies provided by the POI and the risk that the supplies may not available during and after hurricanes. Grocery stores are the main vendors of food and water, which is essential to maintain life needs (Esmalian, Yuan, Xiao, & Mostafavi, 2022). Pharmacies sell prescription medication and first aid supplies, which is especially important for people suffering from chronic diseases. Studies have reported slow recovery of pharmacy operation after hurricanes (Arya, Medina, Scaccia, Mathew, & Starr, 2016; Romolt, Melin, Hardie, Baker, & Louissaint, 2020); Gas station provides fuel for vehicles and generators. The need for gasoline usually could not be wholly satisfied due to the surge of demand and disruptions of supply chains during the onset and post-landfall of hurricanes (Khare, He, & Batta, 2020). The last category is home improvement POIs. People would visit this type of POI to buy backup generators and make construction upgrades for their residence, such as purchasing plywood to cover windows (Chatterjee & Mozumder, 2015). These efforts could reduce the potential damage the hurricane may cause and provide relief for households. This study used NAICS codes to denote the category of POIs. Since each POI is assigned its corresponding NAICS code from the last step, the first four digits were used to filter the four categories from all types of POIs.

### 3.3 Establishing baseline

To detect possible fluctuations in POI visits due to hurricane preparedness activities, we established baselines based on normal period visit patterns. The POI visit data of each CBG in first two weeks of August served as baselines for each category when no disturbance occurred to the residents' normal lifestyle; thus, this period could reflect normal status of residents' POI visit pattern. For each CBG, if the total number of daily visits was smaller than five, that piece of data would not be considered as representative to serve as a baseline. Since the patterns of visits were observed to be different between weekdays and weekends, the baseline period is calculated as a weekly pattern considering each day as a unit.

### 3.4 Calculate percentage change from the baseline

Following the formation and arrival of Hurricane Harvey, preparation activities initiated. Since the hurricane made a landfall on the evening of August 25, 2017, this study defined the preparedness period as days between August 20 and August 25. During the preparation period, each CBG's daily visits to the four categories of POIs were compared to the corresponding baselines to compute the percentage change using Equation 1:

$$P_{C_{i,d,t}} = \frac{E_{i,d,t} - B_{i,d,t}}{B_{i,d,t}} \tag{1}$$

where, $P_{C_{i,d,t}}$ is the percentage change of visits to one category of POI (*t*) from home CBG (*i*)

in date ($d$), $E_{i,d,t}$ is the number of visits to one category of POI ($t$) from home CBG ($i$) on date ($d$). $B_{i,d,t}$ is the calculated baseline value corresponding to the date.

### 3.5 Quantify and visualize metrics of disaster preparedness

This study used two metrics to capture and quantify hurricane preparedness: extent and proactivity. The extent of disaster preparedness was measured by the maximum percentage change of one CBG's visits to each category of POIs during disaster preparation period. Maximum percentage change of POI visits in a CBG means that the number of visits reached a peak compared to normal period; thus, the metric was used to represent the extent of preparedness. Proactivity measures how early the number of visits reached the maximum percentage change for each CBG. Greater proactivity values indicate the residents of the CBG started hurricane preparedness earlier; thus, they had more time to procure necessary supplies ahead of the hurricane. To calculate proactivity, we first recorded the date when maximum percentage for each CBG occurred; then we calculated the difference between the date of maximum percentage change and the date of hurricane landfall. The larger the difference, the earlier the CBG reached the maximum of percentage change in POI visits.

### 3.6 Classify CBGs based on disaster preparedness and evacuation rate

When faced with incoming hurricanes, people choose either to remain shelter in place or to evacuate. A population group becomes highly vulnerable if they chose to shelter in place while unable to adequately prepare for a hurricane. To capture the extent of populations remaining shelter in place, this study took synchronous evacuation data into consideration. The device-level data for evacuation was also obtained from Spectus aggregated datasets. The table Daily Metrics by Device contains users' home census block group, and the hours users remained at their home CBG per day and night. If the location of the device left its home CBG and dwelled in other CBGs for more than 24 hours, the user of the device was counted as evacuated. Evacuation rate was calculated based on the number of evacuated devices divided by the total number of devices in one area. Evacuation rate during normal time (data from July 9 to August 5, 2017) was used as baseline. Comparing evacuation rates between baseline with that during the hurricane preparation period, the percentage change of evacuation rate was calculated to indicate the extent of evacuation.

By dividing preparedness and evacuation rate as high or low level, CBGs of Harris County were classified into four groups. The grouping criterion was based on median value of evacuation rate and extent of preparedness. Accordingly, we identified vulnerability hotspots with low evacuation rate and low preparedness. It is important to note that the evacuation rate is at the census tract level, and the extent of preparedness is at census block group level. To solve the inconsistency, this study assumed that CBGs belonging to the same census tract share the same evacuation rate.

## 4 Results and Findings

### 4.1 Extent of hurricane preparedness

The extent of hurricane preparedness for each CBG was captured by the maximum percentage change of visits to POI compared to the baseline. Based on the extent of disaster preparedness, CBGs in Harris County were classified into three groups. CBGs with a negative value of maximum percentage change were defined as under-prepared areas, because the data indicates that the trend of visit to POIs was decreasing during the preparation period. CBGs whose maximum percentage changes were positive but less than 1 were defined as moderately prepared areas. CBGs whose maximum percentage changes exceeded 1 were defined as highly prepared areas because the trend showed a sharp increase in visiting POIs before a hurricane. Heat maps related to the categorization for four types of POI are shown in Fig. 3.

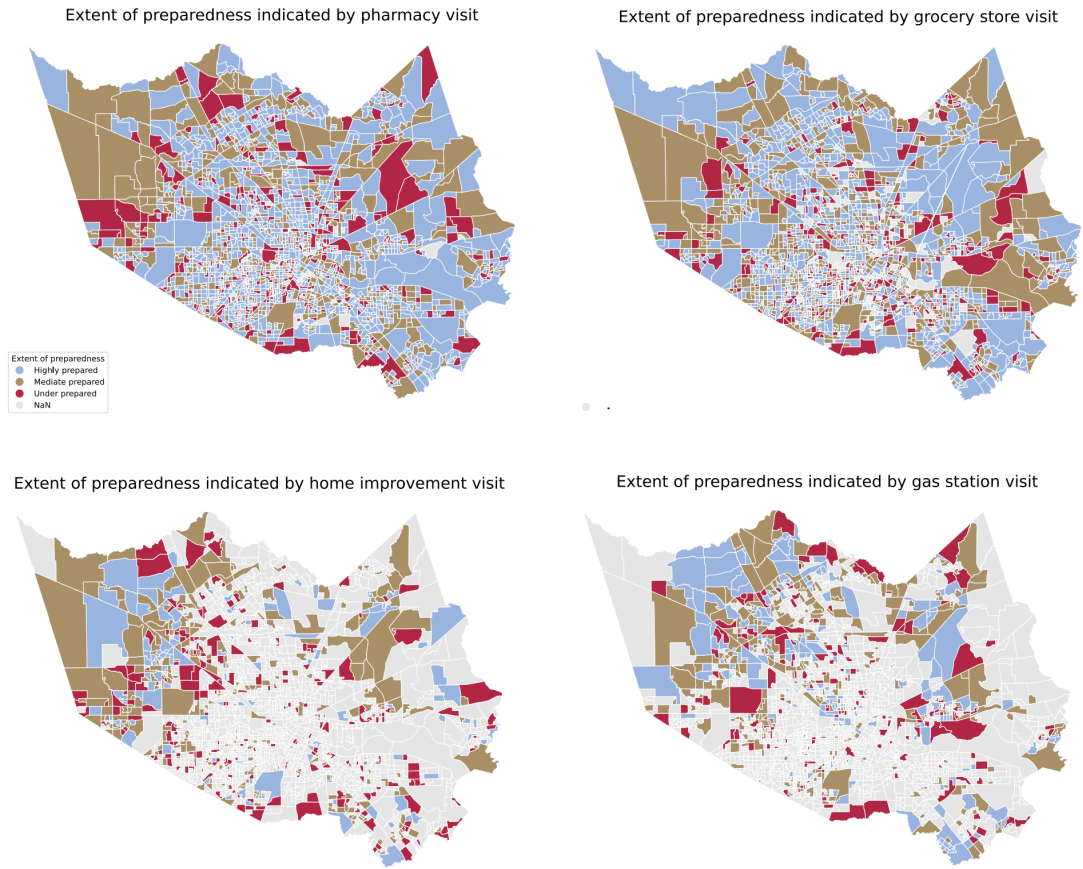

**Fig. 3.** Extent of hurricane preparedness for four types of POI.

Since visits to different types of POIs are highly related to human movements (Zeng et al., 2017), the fluctuation of daily visits before the hurricane could serve as an indicator for hurricane preparedness. For example, the red areas in the maps denote under-prepared CBGs. There is some overlap of red areas across the four subplots in Fig. 3, illustrating that residents in these CBGs reduced their visits to grocery store, pharmacy, gas station and home improvement stores compared to the normal period. Since these POIs are critical facilities for people to maintain normal life and react to disasters, the decrease shows the residents may face the hurricane with insufficient supplies, such as food, water, and medicine. However, there are also variations between the subplots of Fig. 3. For example, the CBGs on the top of the maps show moderate preparedness regarding visits to grocery stores, high preparedness in visiting pharmacies, but low preparedness in visiting gas stations. This phenomenon may indicate residents in these CBGs focused on different aspects of preparing for Hurricane Harvey. The developed metric, extent of preparedness, not only delineates quantitative metrics, but also differentiates the various priorities in households' hurricane preparedness activities. These insights give disaster managers and public officials more complete and detailed information about hurricane preparedness in a proactive manner.

Further analysis was done to explore variations among patterns of visits across the four categories of POIs. Fig. 4 shows the proportion of under-prepared, moderately prepared, and highly prepared groups in the four types of POI. The highest proportion of under-prepared CBGs is in the category of gas stations, while highest proportion of highly prepared CBGs is in the category of visiting pharmacies. That is to say, people prioritized buying necessary medicines and first-aid kits to mitigate the potential impacts of the hurricane but were less concerned with filling vehicle gas tanks before the disaster. Visits to home improvement stores show the second highest rate of under-prepared groups, which may indicate buying hurricane emergency supplies, such as generators and hurricane-resistant windows and doors was neglected by most households.

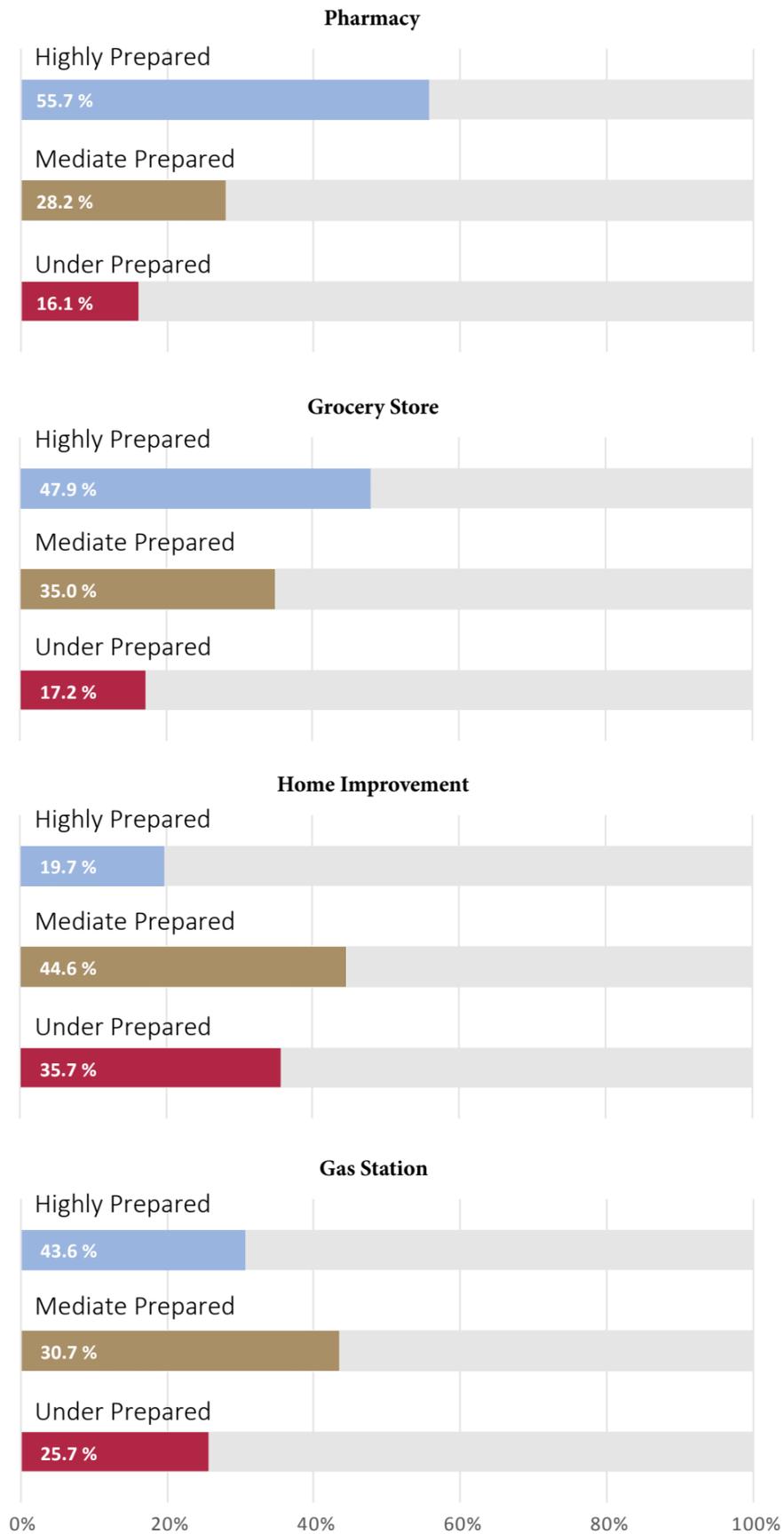

**Fig. 4.** Categories of preparedness extent for four types of POIs.

Fig. 5 shows the distribution of the extent of hurricane preparedness across groups with different income levels. Interestingly, the extent of preparedness is not always higher in groups of higher income. Instead, the data show inconsistent patterns considering visits to different categories of POI. In the gas station category, the results show the trend that the higher income, the higher the extent of preparedness. This is because households with higher income are likely to possess more cars or more likely to have the financial means to evacuate, and thus have more needs to fill their tanks before the hurricane. For pharmacy and home improvement categories though, the results show that there is minor difference between the medium and high-income group, but the low-income group had a greater extent of preparedness. The phenomena may be explained by reasons from two aspects: first, households with low income may face a worse situation even during normal times, compared to other groups. For example, they may not store enough medical supplies at home, and the house they live in may not be kept in a good condition. When encountering an upcoming hurricane, the gaps between their needs and supplies can be exacerbated, which motivated them to devote more efforts in preparation to make up for these disadvantages. Another reason may be that households with low income are more vulnerable to possible impacts of the hurricane: a lower income impedes the protective actions of evacuating or buying homeowner insurance (Ma, Baker, & Smith, 2021) and health insurance (Yamada, Yamada, Chen, & Zeng, 2014), so they will have to pay high prices for medical care or repairs for their houses.

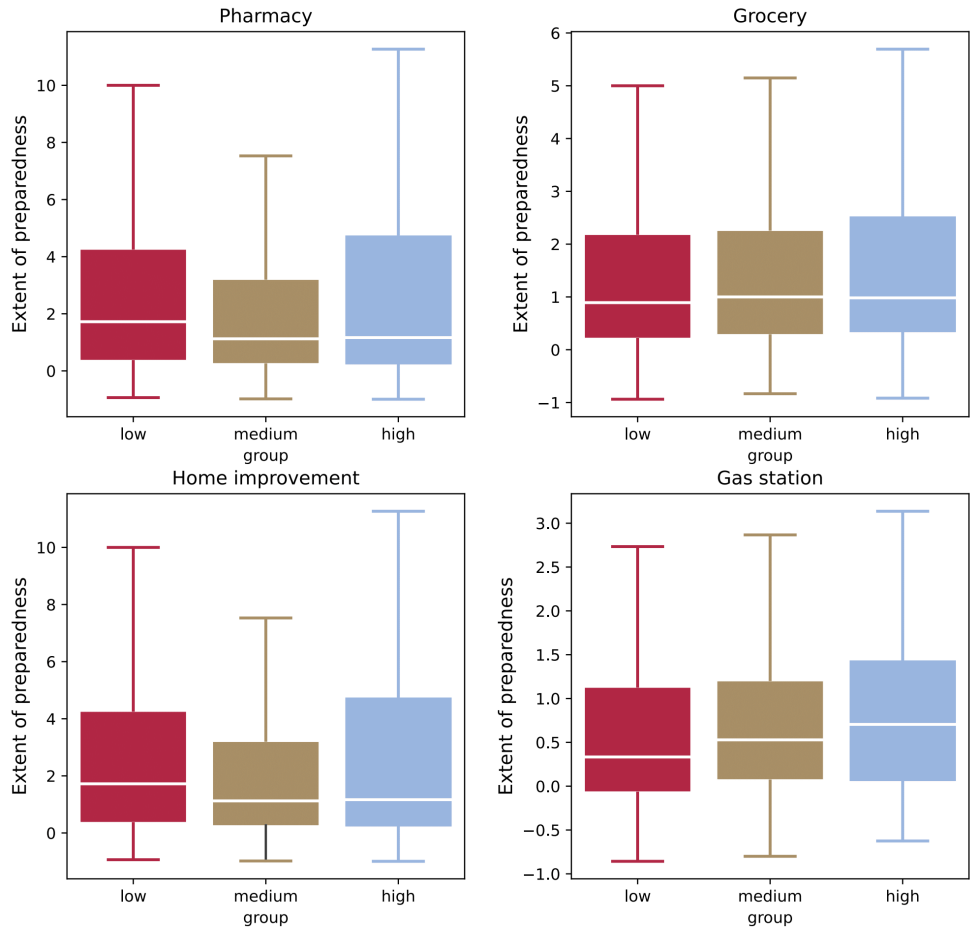

**Fig. 5.** Extent of preparedness measured by maximum percentage change of visits across different groups of income.

### 4.2 Proactivity of disaster preparedness

Proactivity of disaster preparedness captures the timing of hurricane preparedness. This metric was measured by the earliest date when maximum percentage of visits to POI occurred. Fig. 6 shows the percentages of CBGs which have various levels of proactivity. The values in Fig. 6 represent the days between the date when visits to POI reached maximum percentage change and the date of hurricane landfall. The larger the number, the earlier the CBG reached maximum. For example, 0 denotes those CBGs where visits to POIs reached the maximum on August 25.

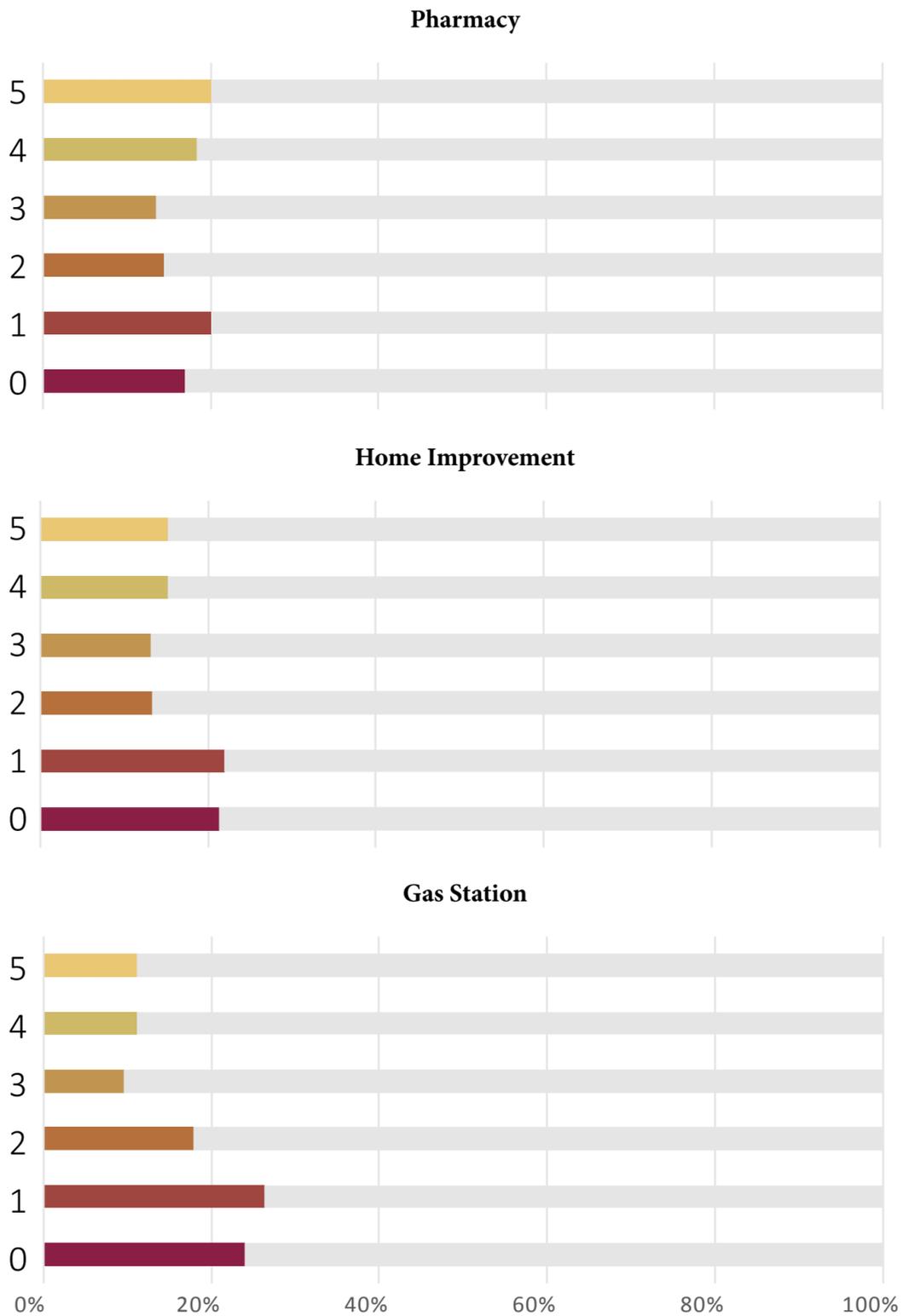

**Fig 6.** Proactivity of CBGs in four categories of POI.

This metric reflects the temporal characteristic of hurricane preparedness. Due to the complex nature of the physics involved in the development of hurricanes, the location, trajectory, and intensity can be uncertain and can change quickly (Chen, Zhang, Carton, & Atlas, 2011). Early preparation could allow ample time for households to purchase supplies. On the other hand, households which start preparation right before the hurricane approaches may face problems, such as not having enough time, or supplies being out of stock. For the four types of POIs, 38.3% of the CBGs made early preparations (shown as bars denoted 5 and 4 in Fig. 6 ) by visiting pharmacies, while the numbers in the other three categories are 30.3% (grocery stores), 22.3% (gas stations), and 30.4% (home improvement stores). The results indicate that most people tended to buy medical supplies in the early period of hurricane preparation. In contrast, the peak of visiting gas stations appears rather late in more than half of the CBGs. Visits to gas station from these CBGs reached maximum percentage change a day earlier or the day before the hurricane made landfall. The phenomena may relate to the characteristics of gasoline consuming: most people rely on vehicles to go to geographically distributed POIs to get supplies, which in turn consumes gasoline. Thus, it's reasonable to assume that people would refill vehicle tanks at gas stations during the late preparation period, thus creating the surge of visits to gas station.

Fig. 7 shows the variance of preparedness proactivity for the four types of POI. From Fig. 7, it can be observed that visits to gas stations show the lowest variance, indicating that residents in CBGs were more consistent regarding the date to fill their tanks before Hurricane Harvey. In addition, the median and quantile are both 1 day, indicating that the visits to gas station in half of the CBGs reached maximum no earlier than 2 days ahead. The metric, proactivity of hurricane preparedness, reveals the temporal characteristics of visits to different categories of POIs during hurricane preparedness period. Since situations where stores may run out of stock of hurricane supplies due to surge of needs occurs from time to time before a disaster (Flood, 2017; Khare et al., 2020), public officials can make use of the metric to track the trend of visits and make adjustments to the hurricane-related supply chain in advance, in order to put sufficient supplies in place for people to prepare for hurricanes. Also, proactivity of hurricane preparedness can be monitored, and the forewarning information can be forwarded precisely to those less active CBGs to remind households to get prepared more actively.

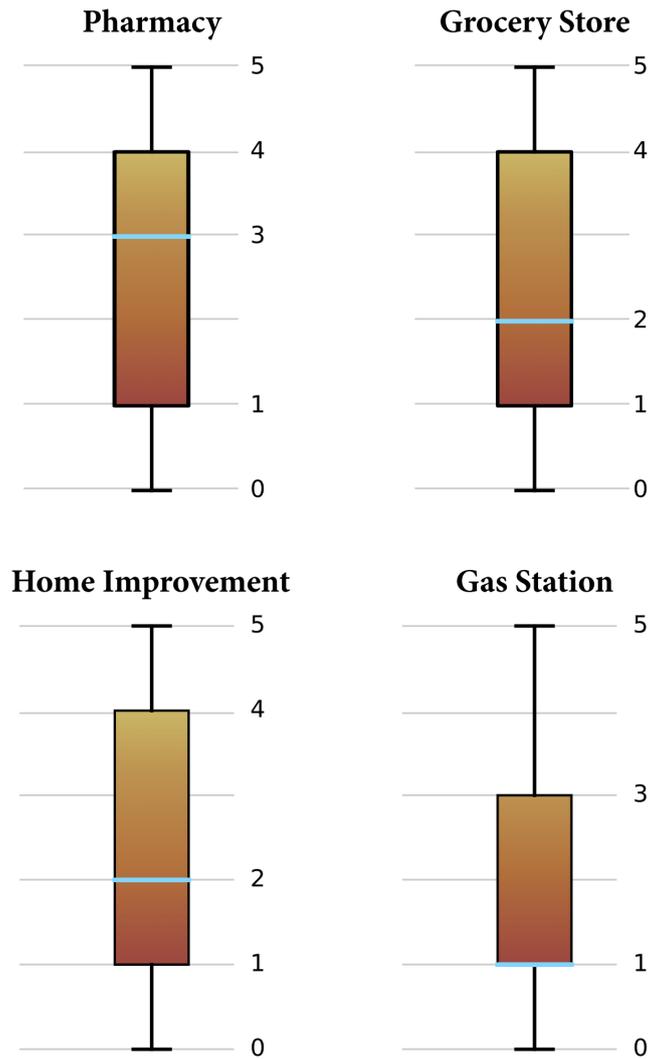

**Figure 7** Boxplot of proactivity for four types of POI

To explore the relationship between the extent of preparedness and proactivity, a correlation analysis was done. Table 2 shows the results of the Spearman correlation analysis including the significance and coefficient for each category of POI. The coefficient between extent and proactivity of hurricane preparedness in terms of visits to grocery stores is not significant, indicating that there is no correlation relationship. For the other three categories of POIs, the correlation between extent and proactivity of disaster preparedness is significant but the correlation is quite weak. The result shows that the extent and proactivity of hurricane preparedness are independent from each other, and the two metrics could reflect unique characteristics of hurricane preparedness.

**Table 2**
Spearman correlation analysis.

| Category of POI | Coefficient | Significance |
|---|---|---|
| Grocery store | 0.018 | 0.417 |
| Pharmacy | 0.081 | 0.000 |
| Gas station | -0.079 | 0.044 |
| Home improvement | -0.155 | 0.000 |

### 4.3 Spatial variation of disaster preparedness and evacuation

This part of the analysis examined the evacuation rate and extent of hurricane preparedness over the same period. CBGs in Harris County were divided into four groups showing the variation of evacuation and hurricane preparedness (Figure 8). Special attention should be given to the overlapping red areas across the four maps, which indicates the CBGs with low evacuation rate and low extent of preparedness considering numbers of visits to all four categories. Most residents in those CBGs neither chose to evacuate nor made enough preparation, thus could be the most vulnerable to hurricane-induced perturbations and impacts. Those shelter-in-place residents may suffer from heavy rainfall, floods, and cascading disruption such as power and telecommunications outage. Without enough supplies, chances are greater that they will experience significant hardship and well-being impacts (Coleman et al., 2020; Dargin et al., 2020; Esmalian, Dong, Coleman, & Mostafavi, 2021). In other words, these areas are the hotspots of susceptibility for emergency management agencies, and they should be prioritized for response and relief supply distribution. By identifying those hotspots in a timely manner, proactive actions could be taken to reduce the impacts of hurricanes on the most vulnerable populations.

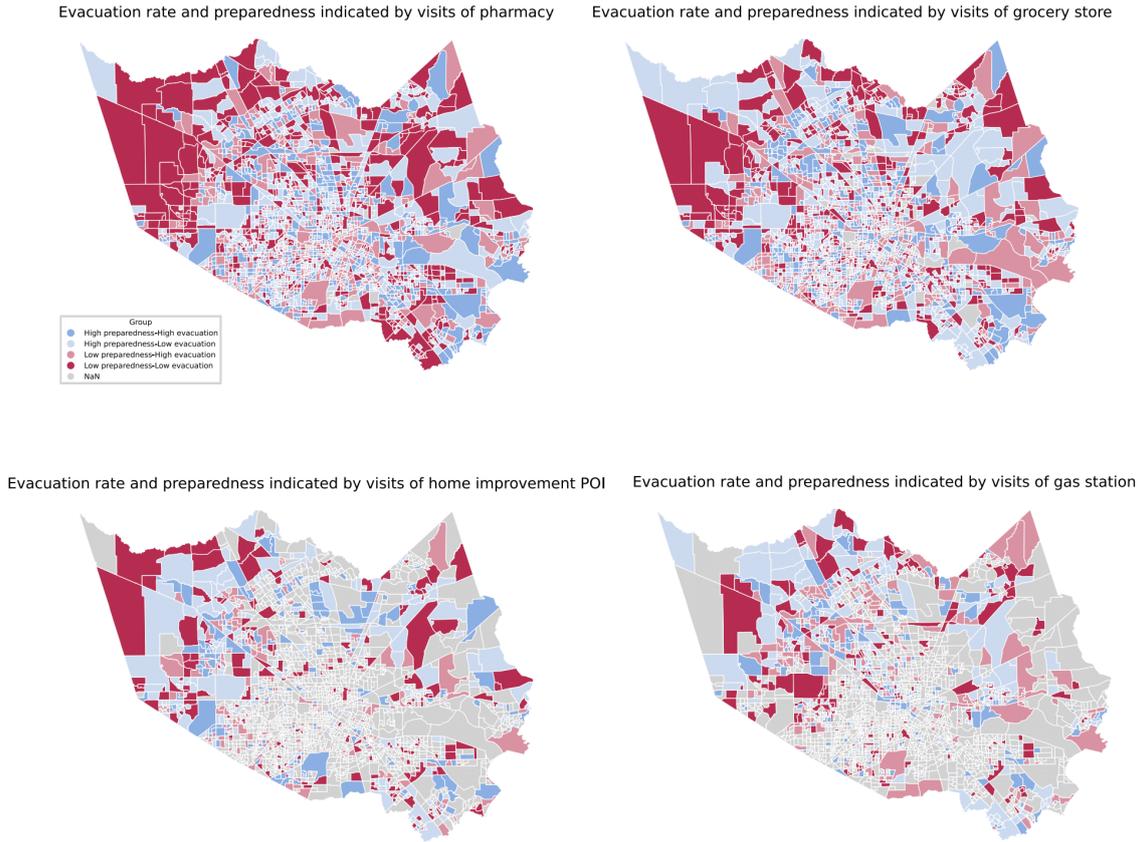

**Fig. 8**. Evacuation and hurricane preparedness maps

## 5   Discussion and concluding remarks

This study used large-scale high-resolution location-intelligence data to capture and quantify the extent, timing, and spatial variations in hurricane preparedness. New metrics were developed and calculated to indicate the extent, timing and spatial characteristics of hurricane preparedness by observing POI visits fluctuation in advance of Hurricane Harvey in 2017. By quantifying visits from each CBG to grocery stores, pharmacies, gas stations and home improvement stores, this study could delineate a clear picture on hurricane preparedness at a fine-grained scale. The analyses of hurricane preparedness metrics showed that among the four categories of POI, the portion of under-prepared CBGs is lowest in the pharmacy category, and highest in gas station category. Regarding the temporal characteristics of disaster preparedness, the visits to pharmacies reached a maximum at the early stage of preparation, while the peak of visits to gas station appeared at the late stage of preparation. The spatial and temporal patterns of visits to grocery store and home improvement store are quite similar, in terms of proportion of preparedness extent groups and proactivity groups. Another important finding is that the extent of hurricane preparedness is independent of proactivity, demonstrating that late preparation is not always accompanied by low extent of preparedness. It inspires us that even preparations are not begun well in advance of a hurricane, households can still achieve a high extent of preparedness before the hurricane. In particular, this study simultaneously considered evacuation rate and the extent of preparedness for CBGs to identify vulnerability hotspots, which were identified by both low preparedness and low evacuation rate.

It is commonly recognized that advance preparation is an essential step to increase resistance and reduce the impacts of a hurricane on households (Kim & Kang, 2010). However, prior research does not offer approaches for quantifying and proactive monitoring of hurricane preparedness at fine spatial scales. The significance of this study is to offer a new data-driven approach and metrics to make use of large-scale location intelligence data to understand hurricane preparedness. By detecting large-scale visits from home CBG to POIs relevant to hurricane preparedness, significant lags in traditional data collection method are overcome, making it possible to understand the characteristics of hurricane preparedness in a timely manner.

This study also offers important insights to emergency managers and public officials. By monitoring the extent and proactivity of human activities related to disaster preparedness, they could examine level of residents' preparedness in every potentially threatened area before the hurricane hits. Then actions such as allocating resources, increasing supplies and distributing reminder information could be taken proactively to improve hurricane preparedness. Based on evacuation rate and extent of preparedness, vulnerability hotspots can be identified and prioritized for relief assistance to reduce impacts on those residents. In addition, the high resolution of data enables precise specification of the CBG location of the hotspots, making the actions more focused and effective. Rather than being reactive and waiting for people to report relief needs, this study allows the emergency managers to response in a proactive, data-driven way.

The limitations of the study are twofold. First, this study mainly adopted location intelligence data collected from smartphones. Since smartphone users tend to be younger or higher-income persons, the elderly and low-income persons are less likely to be included in the data, which may cause some biases (Podesta, Coleman, Esmalian, Yuan, & Mostafavi, 2021). We partially overcome this limitation by using Spectus data, which is shown to have a representative sample of users. Second, from the analysis we could only present how human activities fluctuate before hurricanes but could not discern the reasons for the fluctuations. For example, reasons cannot be clarified via this study for the areas where the visits to POIs decreased during the preparation period. Additional data, such as social media data, should be combined to explore preparedness behaviors in relation with the POI visitation patterns.

**Data availability**

All data were collected through a CCPA- and GDPR-compliant framework and utilized for research purposes. The data that support the findings of this study are available from Spectus, but restrictions apply to the availability of these data, which were used under license for the current study. The data can be accessed upon request submitted on spectus.ai. Other data we use in this study are all publicly available.

**Code availability**

The code that supports the findings of this study is available from the corresponding author upon request.

**Declarations of interest:** none

**Acknowledgement**

This material is based in part upon work supported by the National Science Foundation under Grant CMMI-1846069 (CAREER), Texas A&M University X-Grant 699, and the Microsoft Azure AI for Public Health grant. The authors also would like to acknowledge the data support from Spectus. Any opinions, findings, conclusions or recommendations expressed in this material are those of the authors and do not necessarily reflect the views of the National Science Foundation, Texas A&M University, Microsoft Azure, or Spectus.

Baker, E. J. (2011). Household preparedness for the aftermath of hurricanes in Florida. *Applied Geography, 31*(1), 46-52.

Chatterjee, C., & Mozumder, P. (2015). Hurricane Wilma, utility disruption, and household wellbeing. *International Journal of Disaster Risk Reduction, 14*, 395-402.

Chen, H., Zhang, D.-L., Carton, J., & Atlas, R. (2011). On the rapid intensification of Hurricane Wilma (2005). Part I: Model prediction and structural changes. *Weather and Forecasting, 26*(6), 885-901.

Clay, L. A., & Ross, A. D. (2020). Factors associated with food insecurity following Hurricane Harvey in Texas. *International Journal of Environmental Research and Public Health, 17*(3), 762.

Coleman, N., Esmalian, A., & Mostafavi, A. (2020). Anatomy of susceptibility for shelter-in-place households facing infrastructure service disruptions caused by natural hazards. *International Journal of Disaster Risk Reduction, 50*, 101875.

Comes, T., & Van de Walle, B. A. (2014). Measuring disaster resilience: The impact of hurricane sandy on critical infrastructure systems. *ISCRAM, 11*(May), 195-204.

Dargin, J., Berk, A., & Mostafavi, A. (2020). Assessment of household-level food-energy-water nexus vulnerability during disasters. *SUSTAINABLE CITIES AND SOCIETY, 62*, 102366.

Darzi, A., Frias-Martinez, V., Ghader, S., Younes, H., & Zhang, L. (2021). Constructing Evacuation Evolution Patterns and Decisions Using Mobile Device Location Data: A Case Study of Hurricane Irma. *arXiv preprint arXiv:2102.12600*.

Deng, H., Aldrich, D. P., Danziger, M. M., Gao, J., Phillips, N. E., Cornelius, S. P., & Wang, Q. R. (2021). High-resolution human mobility data reveal race and wealth disparities in disaster evacuation patterns. *Humanities and Social Sciences Communications, 8*(1), 1-8.

Der Sarkissian, R., Cariolet, J.-M., Diab, Y., & Vuillet, M. (2022). Investigating the importance of critical infrastructures' interdependencies during recovery; lessons from Hurricane Irma in Saint-Martin's island. *International Journal of Disaster Risk Reduction, 67*, 102675.

Esmalian, A., Dong, S., Coleman, N., & Mostafavi, A. (2021). Determinants of risk disparity due to infrastructure service losses in disasters: a household service gap model. *Risk analysis, 41*(12), 2336-2355.

Esmalian, A., Yuan, F., Xiao, X., & Mostafavi, A. (2022). Characterizing Equitable Access to Grocery Stores During Disasters Using Location-based Data. *arXiv preprint arXiv:2201.00745*.

Fan, C., Jiang, X., & Mostafavi, A. (2021). Evaluating crisis perturbations on urban mobility using adaptive reinforcement learning. *SUSTAINABLE CITIES AND SOCIETY, 75*, 103367.

farahmand, h., Wang, W., Mostafavi, A., & Maron, M. (2021). Anomalous Human Activity Fluctuations from Digital Trace Data Signal Flood Inundation Status. *Environment and Planning B: Urban Analytics and City Science*, 23998083211069990.

Flood, R. (2017). Hurricane Irma: Panic buying as Florida and Caribbean prepare for storm to hit. Retrieved from https://www.express.co.uk/news/weather/850222/Hurricane-Irma-path-destruction-USA-Florida-panic-buying-storm

Gao, X., Fan, C., Yang, Y., Lee, S., Li, Q., Maron, M., & Mostafavi, A. (2020). Early Indicators of COVID-19 Spread Risk Using Digital Trace Data of Population Activities. *arXiv preprint arXiv:2009.09514*.

Hong, B., Bonczak, B. J., Gupta, A., & Kontokosta, C. E. (2021). Measuring inequality in community resilience to natural disasters using large-scale mobility data. *Nature communications, 12*(1), 1-9.

Josephson, A., Schrank, H., & Marshall, M. (2017). Assessing preparedness of small businesses for hurricane disasters: Analysis of pre-disaster owner, business and location characteristics. *International Journal of Disaster Risk Reduction, 23*, 25-35.
25